# SYMBOLS AND ASTROLOGICAL TERMS IN ANCIENT ARABIC INSCRIPTIONS

**Mohammed H. Talafha[1*] and Ziad A. Talafha[2]**

[1]*Dept. of Astronomy, Eötvös Loránd University, 1117 Budapest, XI. Pázmány Péter sétány 1/A*
[2]*Dept. of History, AL al-BAYT University, 25113 Mafraq, Jordan*

**ABSTRACT**

In the past, the Arabs in Al-hara Zone used many stars to deduce the seasons of the year and also to deduce the roads, at that time this was the most convienent way to figure their ways and to know the time of the year they have to travel or to planet, The most important used stars at that time were the Pleiades, Canopus, Arcturus and other stars. This study shows the inscriptions found in Al-hara Zone in many field trips in the year 2018 which were written on smooth black rocks and how these inscriptions related to the stars and to the seasons – at that time - of the year.

**KEYWORDS:** Arabic, Stars, Inscriptions, Al-hara Zone, Rock art from southern Syria and north-east of Jordan in Badia al-Sham, The Pleiades, Arabian Tribes, Canopus, Seasons, Pre-Islamic era.

## 1. INTRODUCTION

The ancient Arabic inscriptions, which were discovered, were mostly in Al-hara Zone, extends from southern Syria and north-east of Jordan in Badia al-Sham. The map of its spread was from Arar, Badna, northeast Saudi Arabia to Baalbek "Jarad Assal" in Lebanon and Al-Salihiyah in North Lebanon, and from Umm el-Jimal and Azraq in the west of Jordan, to Al-Rutba, in the west of Iraq (Talafha, 2017). These inscriptions were written on smooth black rocks, emerged from the ground as a result of an impact of some volcanoes which spread on the surface, this area contained of many valleys and fertile oases, which was a target for the Arab tribes to settle around such as: the tribe of Amrat, Wrqan, Qeshm, Ḥawlat, Dayef, Yashkur and others, and it is on more than fifty Arab tribes. These tribes lived from the fourth century BC until the third century AD, and they used the northern Arabic musnad line to write on these rocks (Al-Rousan, 1987).

Several studies were made for this particular region and the regions around it, (Gingrich, 2017) studied the rock art from the west and south west Arabia, (Glanzman, 2017) studied Petroglyphs within the Wadi Raghwan in Ma'rib governate in Republic of Yemen, and finally, (Menshawy, 2017) studied the re-consideration and prospective of Qatar cultural heritage tourism map, among other studies.

Many field trips were made to the Jordanian Badia in each of the following areas: Al-Hurra - Wadi Salma, Wadi Al-Fahdah, Wadi Al-Ashaheeb, Ka'ab Al-Abed and Al-Ma'an Hills. Most of the inscriptions found on the rocks of this area were studied taking into account the surfaces erosion that may affect the surface of the rock, the inscriptions containing the writings associated with stars and stellar phenomena were selected and classified according to use at the time, back to the studies and the old literature and some locals (Talafha, 2018).

## 2. METHODOLOGY AND RESULTS

These inscriptions contained mainly about stars that determine the seasons, and the time to change from one season to another, also some of them talk about stars that they used it to know their ways.

### 2.1.1. Seasons

The Arabian tribes used two words to represent the year concept, "Sanah" and "A'am" which both mean year, but the difference is that the first is for years contain good events, and the second for years contained bad events (Al-Ma'anee, 2004), see Figures 1 and 2.

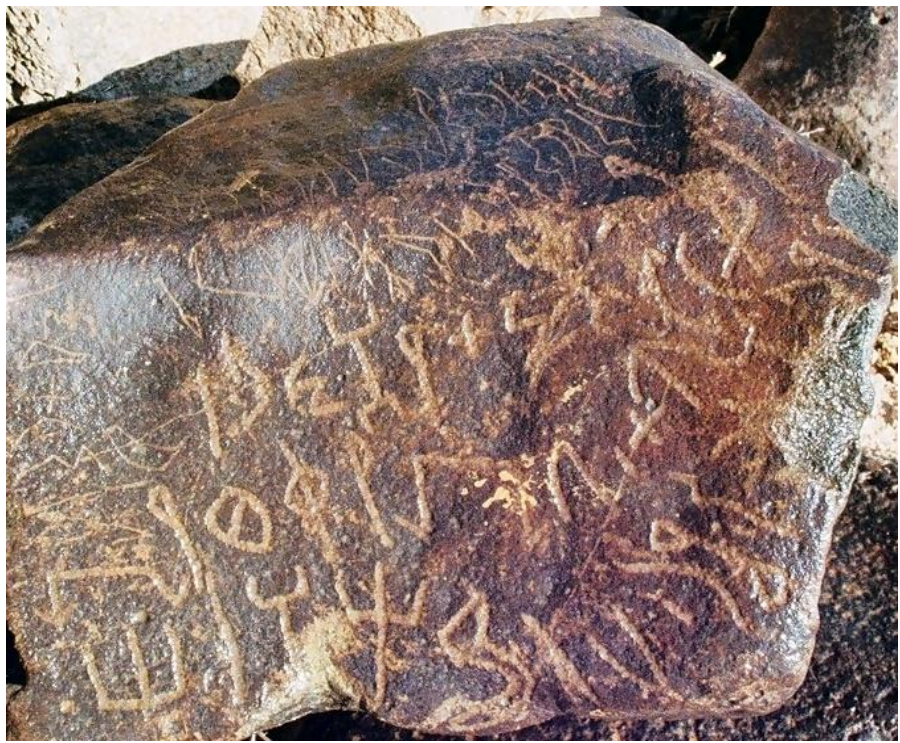

*Figure 1: This inscription contains the words "snt, Sanah" and "wqayẓ" where found as in the sentence " l ġyr'l bn ġt bn slmḏ'lhzywqayẓsnthrbḫlst " (Byġayr'l son ġat son slm of the tribe hzy and he spend the summer in the year the he war ḫalst) where qayẓ is the heat season which come after the spring season and can be inferred from the appearance of "nathra" star and ends by the Arcturus star. (Scale: 5 cm ) (Talafha, 2006)*

| Al - Fahdawi hill | 2007 | 673m | 32.18411 | 037.23878 |

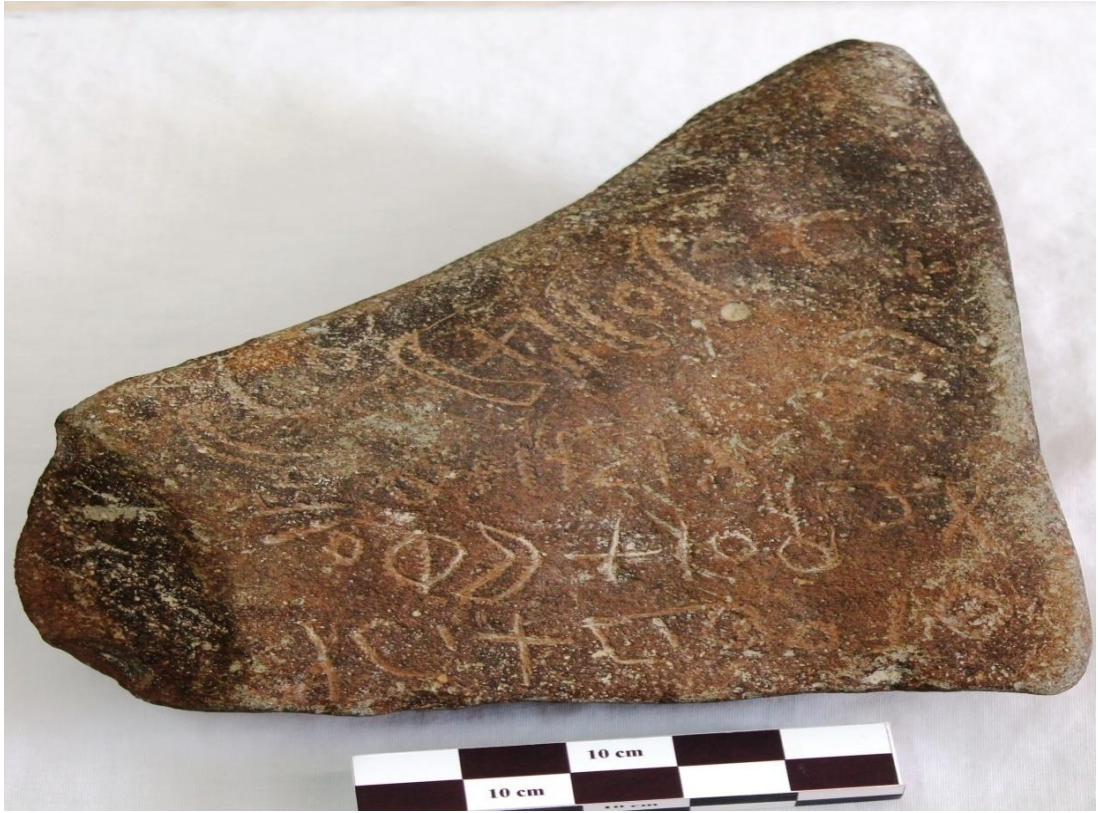

*Figure 2: This inscription contains the word (bʿam ) as follows:*
*"l bnt bn wʾlwḥrsʾl gnm bʿm thnmʾms" (Bybant son waʾlwa-ḥaras on tm and on ganmbʿam tahnmʾams) Bybant son waʾlhe was on the look –out" on tam and on ganm a very hot year (Scale: 10 cm) (Talafha, Z. and Khasawna, A. 2017)*

Then, the terms which points to the seasons was associated to some actions for seasonal traveling these tribes according to certain times, so there life included several travels looking for water and places to stay (Al-Ma'anee, 2004), and in the inscription (Clark, 1984-1985) which was found in Azraq area in the eastern of Jordan, saying "And he took his cattle to graze and went back using the star", this was a usual habit for Arabs to went back to their homes after the end of the grazing season, which was when the Beta Arietis (Sheratan in Arabic), also the Pleiades seen in the sky in the middle of May at that area (Alababneh, 1996).

The Arabian Bedouin used the entering/exiting the heat and cold times, the grow of the grass and when its vanish to determine time among the year (Alababneh, 1996), so for example they known the seasons according to the location of the sun in the sky in one of the moon phases which was divided into 28 phase, starting from the Autumn season, which was defined to Arabs as the rain, see figure 3 and 4, this one come after the end of the heat (qayẓ), see figure 1 and 5, and they didn't use this name as a name of a season rather than a name of the rain, as in the following inscription, "wabaedaha (hukhrfa) al-matar fia allaa tsalam w araei aldaan" (Littmann, 1943), which is the only evidence points to the beginning of the rain in the end of the heat time and the beginning of the cold one, this time is known as the zero season (Sifre) then the rain after it called the "Wasmi" which is the beginning of the spring, see figure 6, at this time the Bedouin start to travel looking for the pastures and rain, this time is marked by the appearance of Alpha Carinae star (Suhiel in Arabic) in the mid of August to the mid of September and the rain come after increased heat named as (Dathe'a), see figure 7, or (Wasmi) where the grass grow after it as in the inscription("wdṯʿ bhbl" "wa-dataṯ' by the Camel WH402 (Winnett & Harding, 1978).

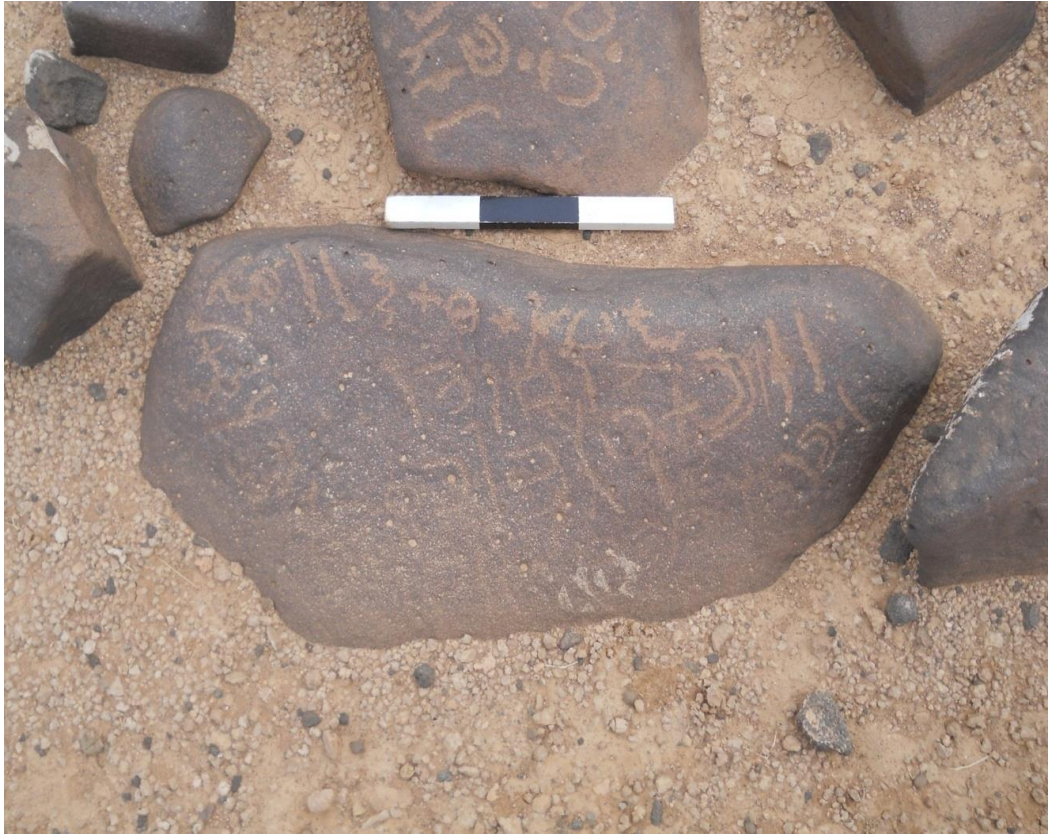

*Figure 3: "L qymt bn aḥrbwwjm hasty"*

*By qaymat son aḥrabwwajamhašaty*
*By qaymat son aḥrab and sadness of the lack of winter*
*(Scale: 5 cm) (Talafha, 2018)*

| Marabhamda | 2018 | 645 m | 32.08167 | 037.27473 |

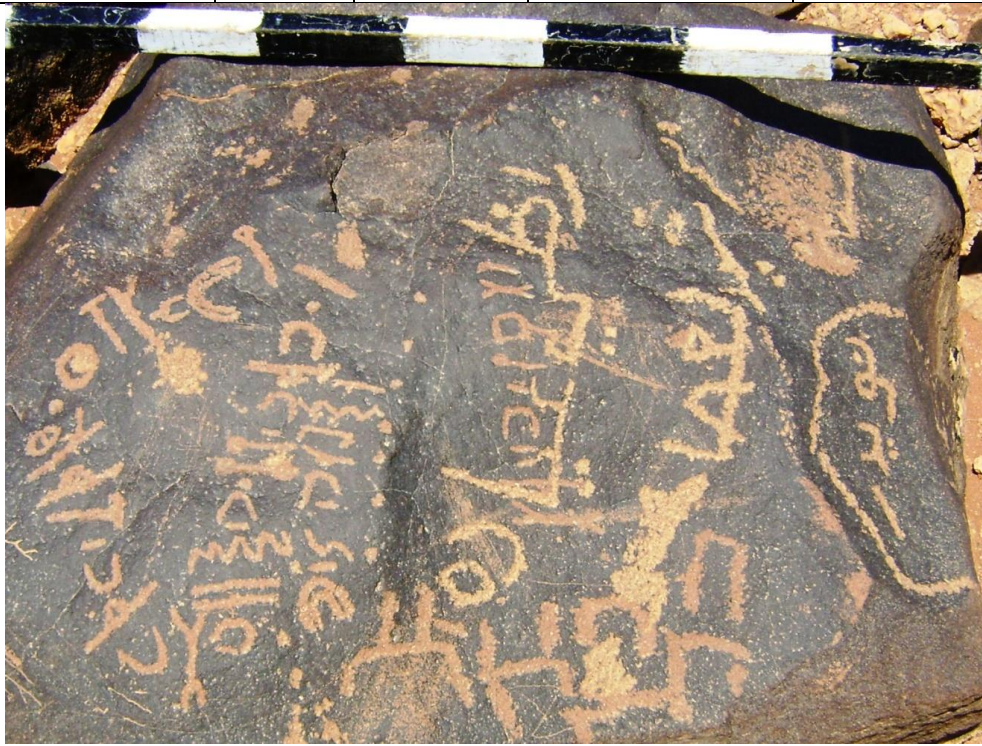

*Figure 4: This inscription saying that the rain didn't come from the sky. (Scale: 5 cm) (Talafha, 2016)*

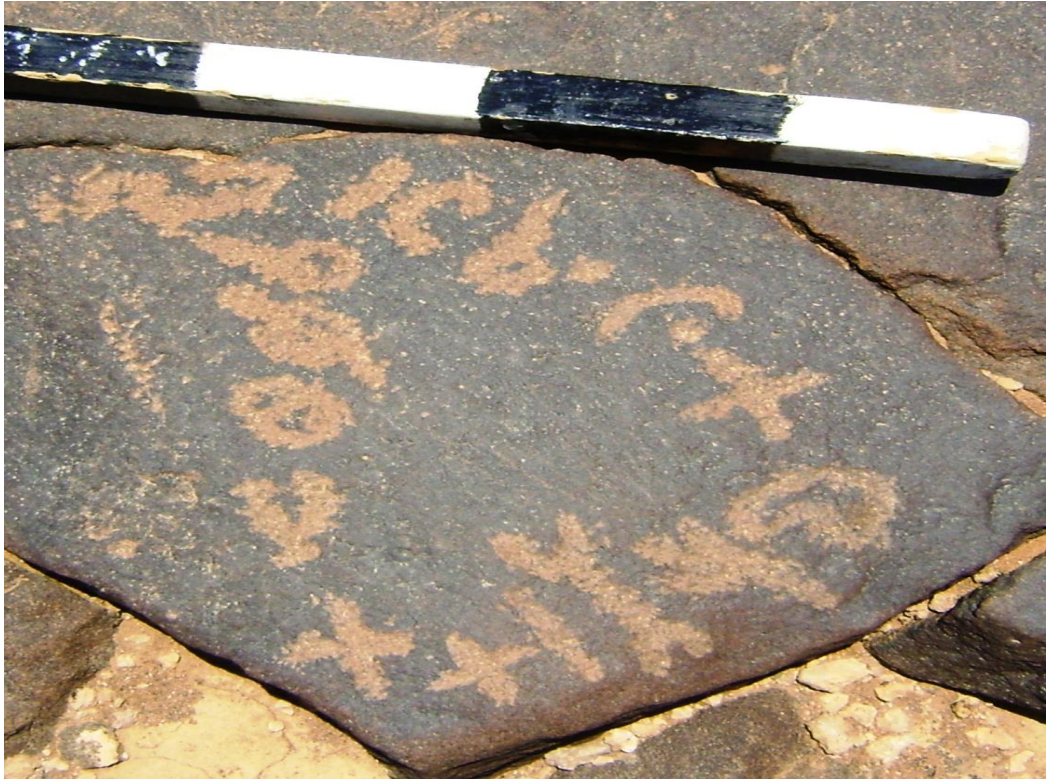

*Figure 5: This inscription contains the word "Qyz"*

**L byn bn tm ḏ'l TtsWqyz**
**By byn son tm of the tribe Tts and he spend the summer (Scale: 5 cm) (Talafha, 2018)**

| Ma'an | 2018 | 648m | 31.99103 | 037.22670 |

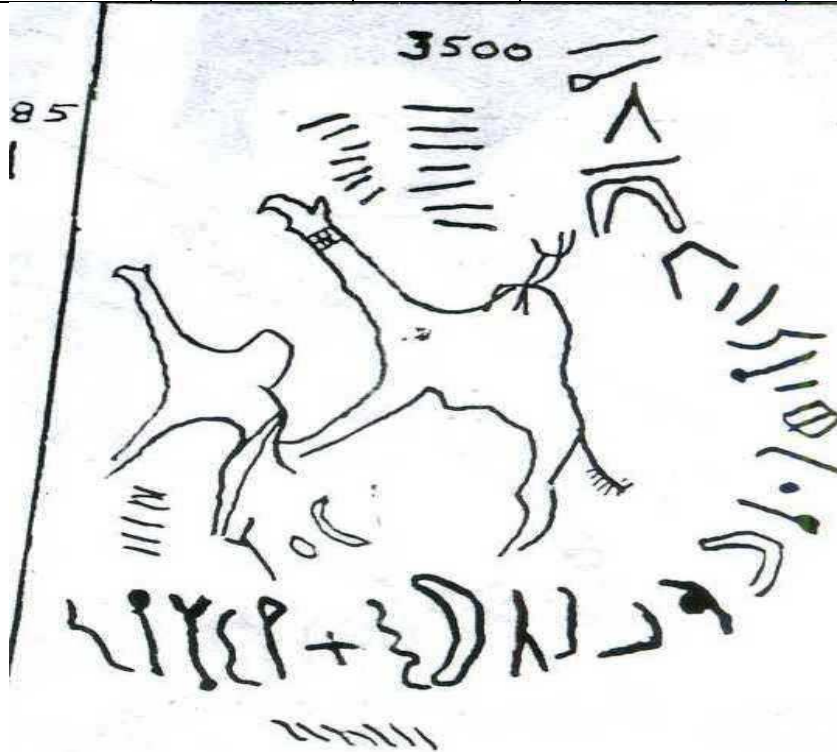

*Figure 6: This inscription contains the word (Saief) "Summer" which come after the winter and before the""Qyz" "qayz" (the spring) (Winnett & Harding, 1978)*

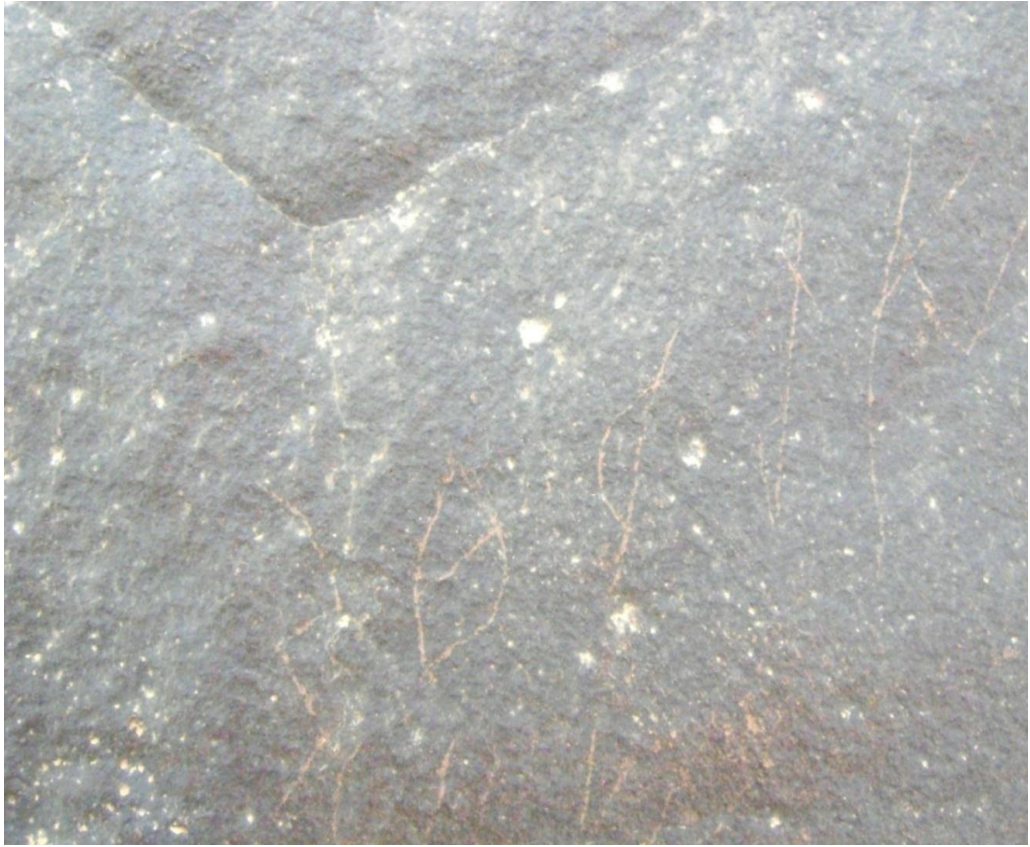

*Figure 7: "wdt'hrḍ""wa-dataṭ' ha" (Scale: 10 cm) (Talafha, 2018)*

| Marabhamda | 2018 | 645 m | 32.08467 | 037.27473 |

Therefore, the seasons according to Arabian tribes are four, (Dathe'a) which is in the end of the Autman season, (Sheti) rain seson in the winter season, (Saief) the summer season which is spring season now and qaied which is the summer season (Alababneh, 1996).

### 2.1.2. Stars

- **Canopus (Suhail)**

The Arabian inscription mentioned the Canopus star as Suhail star as in the following inscription (wnfrhshlsl) "wanafr Suhail hasyl, "the water was soaked by the star Suhail" LP736 (Littmann, 1943).

And it is known that during the appearance of this star, a pouring rain comes which lead to torrent of water, this star is one of the brightest stars in in the southern constellation of Carina, and the second-brightest star in the night-time sky, after Sirius (Alababneh, 1996).

- **Arcturus (Alsamak)**

Arcturus also mentioned as "Alsamak" star which represent the last star in the (Summer) "Saief" season in the following inscription (w n z r s m k m h r n) CIS216 "aintazar samak min huran" (Talafha, 2011).

- **The Pleiades**

The Arabian inscriptions, see figure 8, mentioned the word star (Najem) which may indicate to the Pleiades as some researchers considered (Alababneh, 1996), other saw that it may be refer to the Polaris (Alababneh, 1996), which is the name for the brightest star near the north pole in the northern hemisphere, the ancient Arabs called it "Al-Jade" within the Ursa Minor constellation.

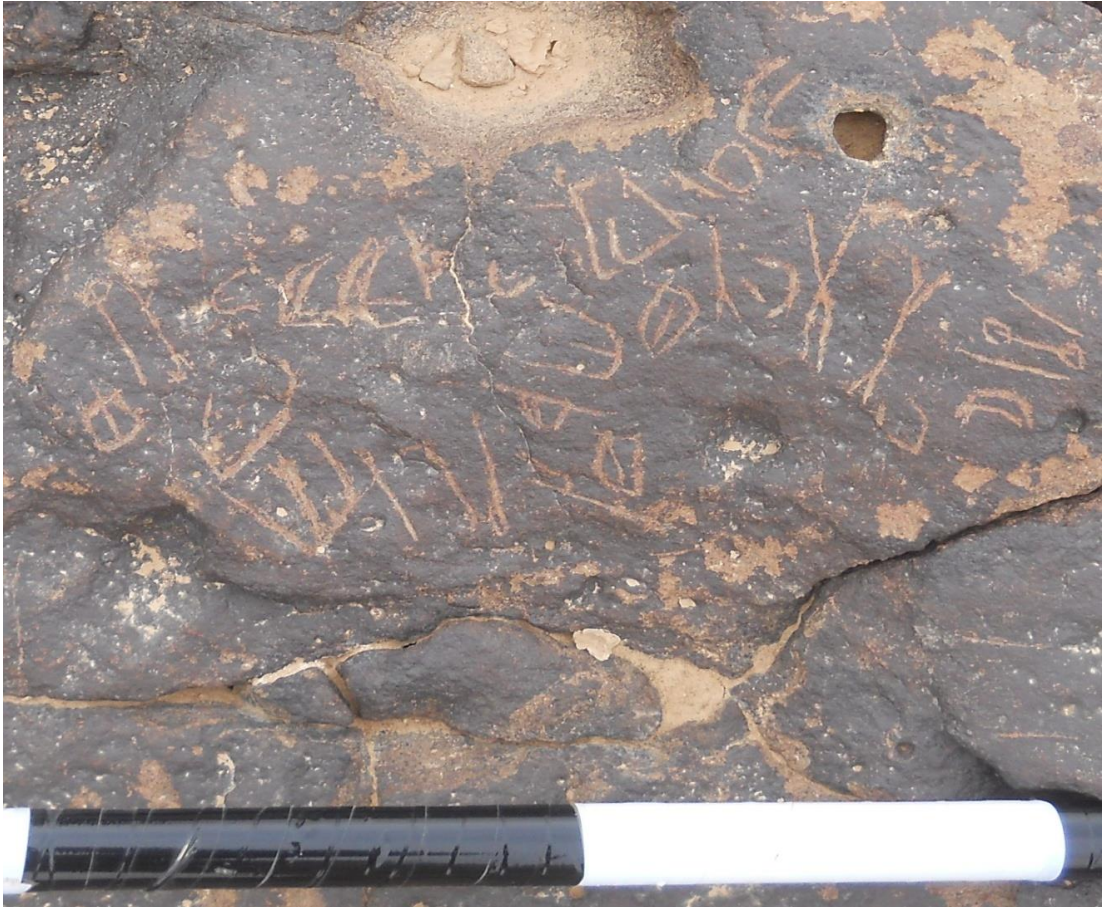

*Figure 8: "lṯlm bn aḫ bn hwḫd waṣir lḫj wyab m mdbr bhnjm".*
*By Thalm son A ibn Khohd and went to Lahj and returned from the desert guided by the star (Scale: 10 cm)*
*(Talafha, 2018)*

Arabs recognized the Polaris by two stars in the Ursae Majoris, Alpha Ursae Majoris and Beta Ursae Majoris and called them "ALdalelan" (Indicators), where the Polaris near by the axis of the earth rotation, and since its near by the horizon one can use it to know the directions, as in the following "lṯlm bn aḫ bn hwḫdwaṣirlḫjwyab m mdbrbhnjm" (Talafha, 2018).

• **Winter hexagon**

In figure 9, one can see six stars and one in the middle, which known in the Arabian culture by the Winter hexagon, these stars are (Capella, Pollux, Procyon, Sirius, Rigel, Aldebaran, Betelgeuse) according to (Alothman, 2018) where he described it in Figs 10, 11.

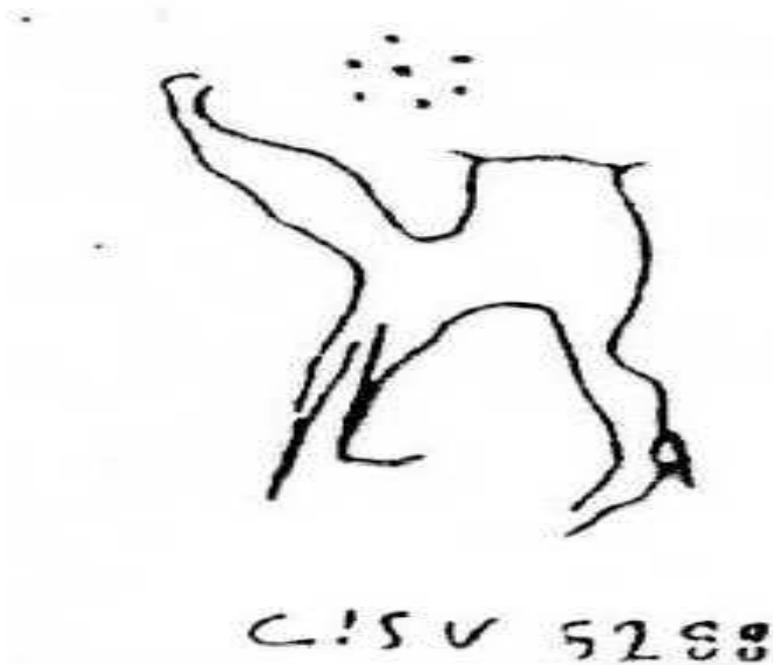

*Figure 9: This figure taken from (Alababneh, 1996) show the claims that the camels get sick in-between the first 10 days of May to the second 10 days of November which is the time that the Pleiades appear in the sky.*

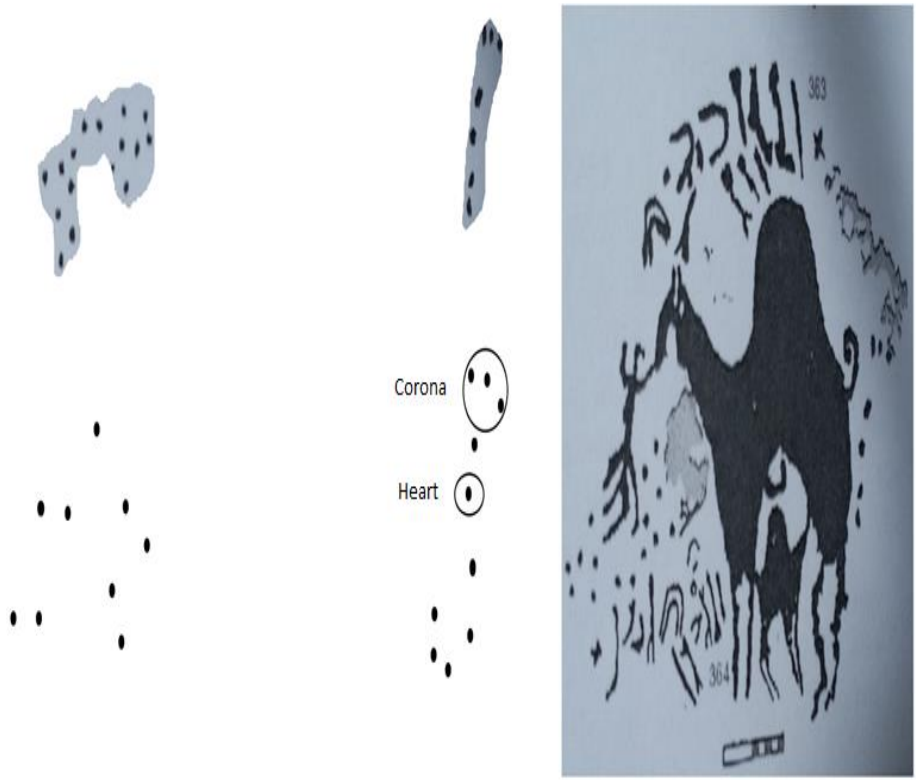

*Figure 10: In the case where the dots in the figure represent the stars, the first combination of these dots which locate behind the camel represent the Scorpius constellations, the second combination represents the Al Warida which represent stars belonging to: Gamma Sagittarii, Delta Sagittarii, Epsilon Sagittarii and Eta Sagittarii(Knobel, E. B. (June 1895). "On a Catalogue of Stars in the Calendarium of Mohammad Al Achsasi Al Mouakket". Monthly Notices of the Royal Astronomical Society. 55 (8): 435. Bibcode:1895MNRAS..55..429K. doi:10.1093/mnras/55.8.429.). (Alothman, 2018)*

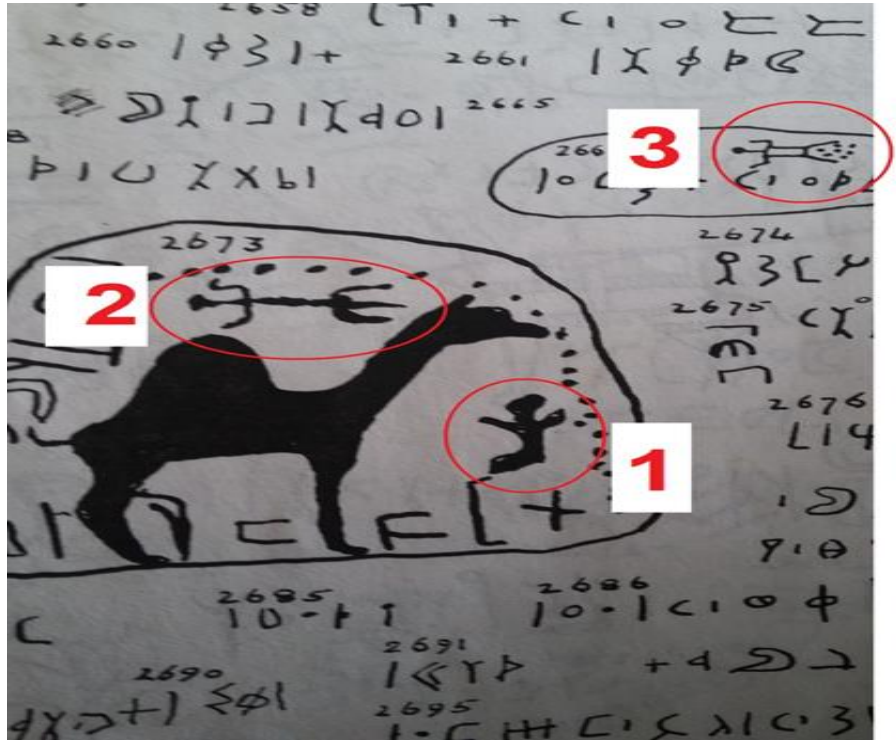

*Figure 11: The star combinations represent: 1) The Pleiades 2) Betelgeuse 3) Leo, caused by the set of the Pleiades, the Betelgeuse appear in the western horizon and the Leo constellation in the middle of the sky. (Alothman, 2018)*

In these excavations, (Talafha, 2018) found another clue as shown in figure 12, where one can see a symbol of a person, with the right hand shorter than the left hand, which was the conception of Arabs for the Pleiades, the right hand was called a leper's hand, and the left one called The Choudib hand which consist of seven stars also we can see it in Fig. 11 from (Alothman, 2018).

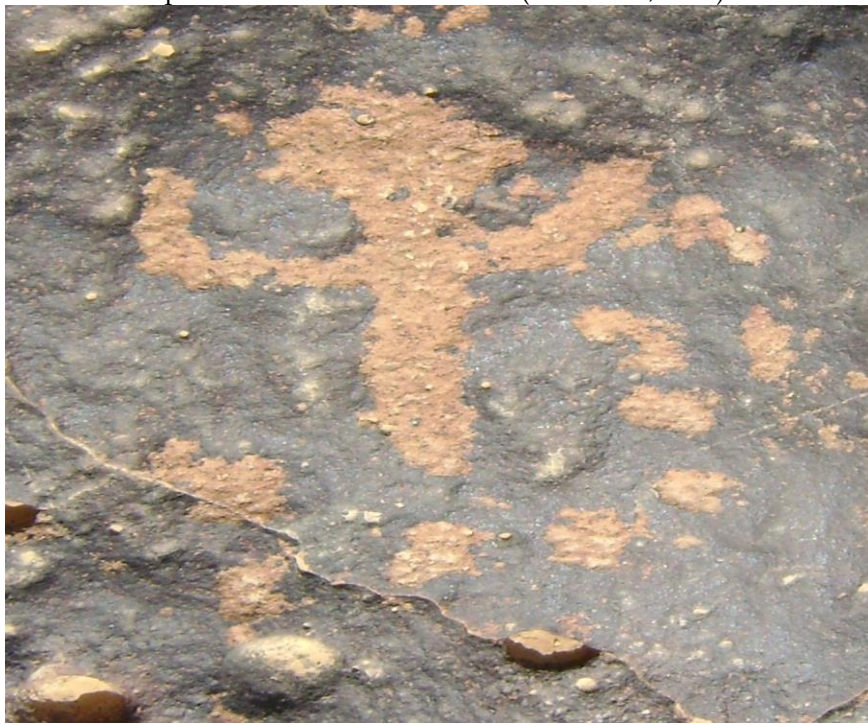

*Figure 12: In this inscription, one can see that the right hand shorter than the left hand, which was the conception of Arabs for the Pleiades, the right hand was called a leper's hand, and the left one called The Choudib hand which consist of seven stars as in the inscription. (Scale: 10 cm) (Talafha, 2018)*

## 3. CONCLUSION

The ancient Arabic inscriptions in the areas studied in this research were presented in two areas. The first was talking about the seasons and their divisions according to the ancient Arabian tribes. The second one was a group of stars that were used at that time and found inscriptions engraved on the rocks of the region like Canopus, Arcturus, The Pleiades and Winter hexagon.